\pdfoutput=1
\def\year{2018}\relax

\documentclass[letterpaper]{article} 
\usepackage{aaai19}  
\usepackage{times}  
\usepackage{helvet}  
\usepackage{courier}  
 \usepackage{url}  
\usepackage{graphicx}  
\usepackage{booktabs} 
\usepackage{multirow}
\usepackage{amsmath}
\usepackage{amsfonts}
\usepackage{enumitem}
\usepackage{xcolor}

\usepackage{footmisc}
\usepackage{todonotes}
\usepackage{subcaption}
\usepackage{colortbl}

\usepackage[normalem]{ulem}
\useunder{\uline}{\ul}{}
\usepackage{caption}

\newcolumntype{C}[1]{>{\centering\arraybackslash}m{#1}}

\title{Analyzing the hate and counter speech accounts on Twitter}
\author{ Binny Mathew$^{1}$ \quad Navish Kumar$^{1}$ \quad Ravina$^{2}$ \vspace{1mm}\\   \Large{\textbf{Pawan Goyal$^{1}$}} \quad  \Large{\textbf{Animesh Mukherjee$^{1}$}}\\
	$^1$ Indian Institute of Technology, Kharagpur \\
	$^2$ National Institute of Technology, Patna\\
	{\tt \small binnymathew@iitkgp.ac.in, navish.kumar27@gmail.com, ravina01997@gmail.com} \\
	{\tt \small pawang.iitk@gmail.com , animeshm@gmail.com}
}

\begin{document}
	
\maketitle

\begin{abstract}
	
The online hate speech is proliferating with several organization and countries implementing laws to ban such harmful speech. While these restrictions might reduce the amount of such hateful content, it does so by restricting freedom of speech. Thus, an promising alternative supported by several organizations is to counter such hate speech with more speech.
In this paper, We analyze hate speech and the corresponding counters (aka counterspeech) on Twitter. We perform several lexical, linguistic and psycholinguistic analysis on these user accounts and obverse that counter speakers employ several strategies depending on the target community. The hateful accounts express more negative sentiments and are more profane. We also find that the hate tweets by verified accounts have much more virality as compared to a tweet by a non-verified account. While the hate users seem to use words more about \textit{envy, hate, negative emotion, swearing terms, ugliness}, the counter users use more words related to \textit{government, law, leader}. We also build a supervised model for classifying the hateful and counterspeech accounts on Twitter and obtain an F-score of 0.77. We also make our dataset public to help advance the research on hate speech.

\end{abstract}

\section{Introduction}

The online proliferation of hatespeech\footnote{https://goo.gl/4rWGiF} has caused several countries and companies to implement laws against hatespeech to enforce the citizens to restrain from such behavior. Countries like Germany, United States of America, France have laws banning hatespeech. Social media sites such as Twitter, Facebook usually respond to hatespeech with the suspension or deletion of the message or the user account itself.

While these laws may reduce the amount of hatespeech in online media, it does so at the cost of causing harm to the freedom of speech. Another potential alternative to tackle hatespeech is ``counterspeech''. The main idea behind counterspeech is to add more speech to the conversation and try to change the mindset of the hate speaker.

This requirement led countries and organizations to consider counterspeech as an alternative to blocking~\cite{gagliardone2015countering}. The idea that `more speech' is a remedy for harmful speech has been familiar in liberal democratic thought at least since the U.S. Supreme Court Justice Louis Brandeis declared it in 1927. There are several initiatives with the aim of using counterspeech to tackle hatespeech. For example, the Council of Europe supports an initiative called `No Hate Speech Movement'\footnote{\label{nohatespeechmovement}No Hate Speech Movement Campaign:  \url{http://www.nohatespeechmovement.org/}} with the aim to reduce the levels of acceptance of hatespeech and develop online youth participation and citizenship, including in Internet governance processes. UNESCO released a study~\cite{gagliardone2015countering} titled `Countering Online Hate Speech', to help countries deal with this problem. Social platforms like Facebook have started counterspeech programs to tackle hatespeech\footnote{\label{counterfb} Counterspeech Campaign by Facebook: \url{https://counterspeech.fb.com/en/}}. Facebook has even publicly stated that it believes counterspeech is not only potentially more effective, but also more likely to succeed in the long run ~\cite{bartlett2015counter}. Combating hatespeech in this way has some advantages: it is faster, more flexible and responsive, capable of dealing with extremism from anywhere and in any language and it does not form a barrier against the principle of free and open public space for debate.

\subsection{The present work}
In this paper, we perform the first comparative study of the interaction dynamics of the hatespeech and the corresponding counterspeech replies. We choose Twitter as our source of data and curate a dataset with 1290 hate tweet and counterspeech reply pairs. After the annotation process, the dataset consists of 558 unique hate tweets from 548 user and 1290 counterspeech replies from 1239 users. We found that 75.39\% of these replies are counterspeech that oppose the hatespeech spread by the user.

\subsection{Contributions}
The main contributions of our paper are as follows:
\begin{itemize}
	\item We perform the first comparative study that looks into the characteristics of the hateful and counter accounts.
	
	\item We provide a dataset of 1290 tweet-reply pair in which the tweets are the hatespeech and their replies are counterspeech. We plan to release the dataset upon acceptance.
	
	
	\item We develop a model which predicts if a given Twitter user is a hateful or counterspeech account with an accuracy of 78\%.
	
	
	
\end{itemize}

\subsection{Observations}
Our study results in several important observations. 
\begin{itemize}
	\item First, we find significant difference in the activity pattern between these users. Hateful accounts tend to express more negative sentiment and profanity in general. If the hateful tweet is from a verified account, it seems to be much more viral as compared to other hateful tweets. 
	\item Another intriguing finding is that hateful users also act as counterspeech users in some situations. In our dataset, such users use hostile language as a counterspeech measure 55\% of the times. 
	\item In terms of personality traits, the counterspeakers seem to have a higher quotient of `agreeableness'. They are more altruistic, modest and sympathetic. The hateful users, on the other hand, seem to have a higher quotient of `extraversion' indicating that they are more energetic and talkative in nature. They are more cheerful, excitement-seeking, outgoing, and sociable in nature. 
	\item Using the linguistic structure of the general tweets and the account characteristics of the hateful and the counter speakers, it is possible to distinguish them early with an accuracy of 78\%.  
	\item We analyze the topical inclination of the users and observe that counterspeakers prefer topics related to `politics', `news', and `journalism' while hate users prefer topics like `niggas'.
\end{itemize}

\section{Preliminaries}

\begin{itemize}
	
	\item \textbf{Hatespeech:} We define hatespeech according to the Twitter guidelines. Any tweet that `promotes violence against other people on the basis of race, ethnicity, national origin, sexual orientation, gender, gender identity, religious affiliation, age, disability, or serious disease' is considered as a hatespeech\footnote{\label{twitter_hate_policy}https://help.twitter.com/en/rules-and-policies/hateful-conduct-policy}. A \textbf{hate account (HA)}
	is a Twitter account which posts \textit{one or more} such hateful tweets.
	
	\item \textbf{Counterspeech:} We follow the definition of counterspeech used in ~\cite{mathew2018thou}. We call a tweet as a `counterspeech' if the tweet is a direct reply to a hateful tweet. A \textbf{counter account (CA)} is a Twitter account which posts one or more such counterspeech in response to a hateful post.
	
\end{itemize}

\section{Dataset Collection}

We use Twitter as the platform for this study. As we are interested in the replies received by a particular hate tweet, we could not use the Twitter API directly as it does not provide any service that helps in directly collecting the replies of a particular tweet. To collect these replies we utilize the PHEME script\footnote{\label{pheme_script}https://github.com/azubiaga/pheme-twitter-conversation-collection} which  allows us to collect the set of tweets replying to a specific tweet, forming a conversation. Other works~\cite{zubiaga_rumours_2016} have used this technique as well. Our overall methodology is a multi-step approach for creating the dataset which can be divided into the following three steps:

\begin{itemize}
	\item Hateful tweet collection.
	\item Filtration and annotation of the hateful tweets.
	\item Extraction and annotation of the reply tweets.
\end{itemize}

We explain these three steps in detail below.





\subsection{Hatespeech collection}
We rely on the hatespeech templates defined in~\cite{silva2016analyzing} to collect the hateful tweets. These templates are of the form:

\begin{center}
	
	I \textless intensity\textgreater \textless userintent\textgreater \textless hatetarget\textgreater
	
\end{center}

The subject ``I'' implies that the user is expressing her own personal emotions. The verb, embodied by \textless userintent\textgreater \hspace{0.5mm} component is used to specify the user's intent which in this case is the word `hate' or its synonyms. The component \textless intensity\textgreater~is optional and acts as a qualifier which some users use to amplify their emotions. Words such as `really',`f**cking' are used to express the  \textless intensity\textgreater. In our work, we have used the words that are listed in ~\cite{mondal2017} for each of the component. The component  \textless hatetarget\textgreater~is used to find the target community on the receiving end of the hatred. Table~\ref{tab:community_keywords} lists the keywords that we have used to extract hateful tweets for each of these communities. Some examples of the these templates include: ``I hate muslims'', ``I really despise white people'' etc.

Next, we utilize the Jefferson-Henrique's web scraping script\footnote{https://github.com/Jefferson-Henrique/GetOldTweets-python} to collect the hateful posts using the templates defined above. We run the script for all the communities defined in Table~\ref{tab:community_keywords} to collect around 578K tweets in total. We removed all non-English tweets from the dataset.

\begin{table*}[!ht]
	\resizebox{\textwidth}{!}{
		
		\begin{tabular}{ p{4cm}  p{8cm}  p{8cm} | C{1cm} } 
			\hline 
			\textbf{Counterspeech Type} &	\textbf{Hatespeech} &	\textbf{Counterspeech} & \textbf{Total}  \\ \hline \hline
			
			Presentation of facts	&	A tragedy that is a DIRECT result of allowing too many f**king Muslims in their country. There should not be any surprises here. \#BANMUSLIMS 	&	The guy was born in UK in 1995 & 136  \\ \hline
			
			Pointing our hypocrisy	&	I hate Fat ppl bruhhhh ughhh	&	``@user: I hate Fat ppl bruhhhh ughhh'' you use to be fat you hypocrite!	&	177		\\ \hline
			
			Warning of consequences	&	I hate lesbians so much omfg	&	I report you to the LGBT community, this blatant discrimination will not be tolerated	&	69	\\ \hline
			
			Affiliation				&	I'm christian so i hate muslims, muslim turkish people and MUSLIM TURKISH BELIEBERS cause YOU ARE THE TERRORIST!Turkish BELIEBERS Need Bieber	&	@user IM MUSLIM BUT İ DO NOT HATE CRISTIANS	&	58		\\ \hline
			
			Denouncing speech		&	I hate gays		&	``@user: I hate gays.'' Stop being homophobic	&	132	\\ \hline
			
			Images					&   - - - 	&   - - - 	&	83	 \\	\hline
			
			Humor					&	I hate gays	&	@user so uhhh I guess you hate yourself huh? HAHAHHAHAHAHAHA man I'm funny	&	134	 \\ \hline
			
			Positive tone			&	@user the problem with the world is you f**king muslims! go choke  to death on some bacon you child raper pig! f**k you a**hole!	&	@user, I'm sorry you feel that way. Islam is a beautiful religion, so misunderstood. Not all Muslims are the same.	&	175	\\ \hline
			
			Hostile language		&	@user f**k the muslims!!!	&	@user you are truly one stupid backwards thinking mother**cker to believe negativity about Islam	&	357	 \\ \hline
			
			Miscellaneous		&	F**k I hate women! All bit**es to a good guy	&	@user Hey hey now.. ALL women are NOT the same. 	&	220	 \\ \hline \hline
			\multicolumn{3}{c}{\textbf{Total}} & 1481\\
			
			\hline
	\end{tabular}}
	\caption{Example hatespeech and the corresponding counterspeech received for each type. Please note that the total is more than the reported 1290 counterspeech replies because the users sometimes employ multiple strategies in a single tweet. }
	~\label{tab:counterspeech_examples}
\end{table*}

\begin{table}[ht]
	\begin{tabular}{l l l}
		{\ul \textbf{Gender}}             & {\ul \textbf{Ethnicity}}   & {\ul \textbf{Physical Trait}} \\
		Men                               & Nigger                     & Fat                                                                   \\
		Women                             & Nigga               &                                                                       \\
		Female                            & White people               &                                                                       \\ 
		& Black people                            &                                                                       \\
		&                         &                                                                           \\
		{\ul \textbf{Sexual Orientation}} & {\ul \textbf{Nationality}} & {\ul \textbf{Religion}}                                               \\
		Lesbian                           & American                   & Jew                                                                   \\
		Gay                               & Indian                     & Islam                                                                 \\
		LGBT                              & Mexican                    & Muslim                                                                \\
		Transgender                       & Arab                       &                                                                       \\
		& Asian                      &                                                                      
	\end{tabular}
	\caption{Keywords used during search to select hatespeech for different communities.}
	~\label{tab:community_keywords}
\end{table}

\begin{figure}[t]
	\centering
	\includegraphics[width=.5\textwidth]{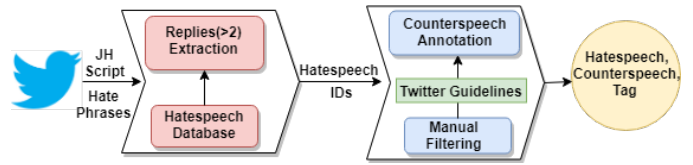} 
	\caption{A flowchart showing the process by which we obtained the dataset.}
	~\label{fig:dataset_flowchart}    
\end{figure}

\subsection{Filtration and annotation of hateful tweets}
As an initial filtering step to ensure that we get sufficient replies to a hateful tweet, we filter out all the tweets collected in the previous step that did not have at least two replies. This reduced the number of hatespeech tweets to 13,321. Next, we manually tagged the tweets as containing hatespeech or not. We follow the guidelines defined by Twitter~\footref{twitter_hate_policy} to identify if a tweet contains hatespeech or not. Each tweet was annotated by two users and in case of a disagreement, we ask one more annotator to resolve it. The annotators were instructed to take the context into consideration when annotating a tweet as hateful. The annotations were performed by a group of two undergraduate students with Major in Computer Science and a PhD student in Social Computing. The final dataset consisted of 558 tweets that were tagged as hateful. Figure~\ref{fig:dataset_flowchart} illustrates the whole process in a flowchart.

\subsection{Extraction and annotation of the reply tweets}
Once we selected the hatespeech tweets with at least two replies, the next step involved using the PHEME script\footref{pheme_script} to scrape the replies received to these hateful tweets. We observe that the 558 hate tweets received a total of 1711 replies with a mean of 3.067 and a median of 2 replies per tweet. We employ two annotators for the counterspeech tagging. We follow the procedure used in ~\cite{mathew2018thou} to annotate the counterspeech data. For each reply text to a hatespeech tweet, the annotators were asked to perform two tasks: (i) decide if the reply is a counterspeech or not, and (ii) if the reply is indeed a counterspeech, then decide the category of the counterspeech. We describe these categories in the subsequent section.

Two independent annotators tagged each reply tweet as counterspeech or not and got an accuracy of 81.35\% and $\kappa$ score of  0.46, which is moderately good. For the second task of deciding on the category of counterspeech, we use the Exact Match Ratio and Accuracy defined in~\cite{sorower2010literature}. The two annotators obtain an Exact Match Ratio of 71.4\% and an Accuracy of 77.58\%. We employ a third annotator to resolve the conflicting cases and report the final distribution in Table~\ref{tab:counterspeech_examples}.

\section{Taxonomy of Counterspeech}

A successful counter to a hatespeech will require various types of strategies. In~\cite{susan2016counterspeech}, the authors define eight such strategies that are used by counterspeakers. Similar to~\cite{mathew2018thou}, we divide the \textit{Tone} category into two parts: positive  and negative Tone.  We aggregate all the counterspeech that would not fit in the above categories into a `Miscellaneous' category. Note that sometimes the users employ multiple counterspeech strategies in a single tweet. We refer the readers to~\cite{susan2016counterspeech,mathew2018thou} for a detailed discussion on the taxonomy of counterspeech. Table~\ref{tab:counterspeech_examples} shows an example for each type of counterspeech.



\begin{table*}[!ht]
	\resizebox{\textwidth}{!}{
		\begin{tabular}{l*{7}{r}} 
			\hline 
			\textbf{Hate Target} &	\textbf{Gender} &	\textbf{Sexuality} & \textbf{Nationality}&	\textbf{Religion} & \textbf{Physical Trait}&	\textbf{Ethinicity} & \textbf{Total}\\ \hline \hline
			
			Presentation of facts   & 1 (00.36\%)	&	5 (02.54\%)	    &	5 (04.24\%)	        &\cellcolor{green}125 (17.86\%)	&	0 (00.00\%)	        &	2 (00.96\%)	    &	138 (08.39\%)\\
			Pointing out hypocrisy  & 38 (13.77\%)	&	19 (9.64\%)	    &	16 (13.56\%)	    &	\cellcolor{green}104 (14.86\%)	&	7 (4.86\%)	        &	7 (3.35\%)	    &	191 (11.62\%)\\
			Warning of consequences & 3 (01.09\%)	&	9 (4.57\%)	    &	4 (3.39\%)	        &	35 (5.00\%)	    &	2 (1.39\%)	        &	\cellcolor{green}25 (11.96\%)	&	78 (4.74\%)\\
			Affiliation             & 14 (05.07\%)	&	9 (4.57\%)	    &	\cellcolor{green}9 (7.63\%)	        &	24 (3.43\%)	    &	2 (1.39\%)	        &	4 (1.91\%)	    &	62 (3.77\%)\\
			Denouncing speech       & 15 (05.43\%)	&	20 (10.15\%)	&	12 (10.17\%)	    &	53 (7.57\%)	    &	3 (2.08\%)	        &	\cellcolor{green}34 (16.27\%)	&	137 (8.33\%)\\
			Images                  & 17 (06.16\%)	&	10 (5.08\%) 	&	\cellcolor{green}10 (8.47\%)	        &	41 (5.86\%)	    &	1 (0.69\%)	        &	10 (4.78\%)	    &	89 (5.41\%)\\
			Humor                   & 32 (11.59\%)	&	\cellcolor{green}30 (15.23\%)	&	6 (5.08\%)	        &	51 (7.29\%)	    &	12 (8.33\%)	        &	8 (3.83\%)	    &	139 (8.45\%)\\
			Positive tone           & 47 (17.03\%)	&	\cellcolor{green}\cellcolor{green}34 (17.26\%)	&	13 (11.02\%)    	&	64 (9.14\%)	    &	15 (10.42\%)	    &	13 (6.22\%)	    &	186 (11.31\%)\\
			Hostile language        & 50 (18.12\%)	&	39 (19.80\%)	&	32 (27.12\%)    	&	124 (17.71\%)	&	\cellcolor{green}65 (45.14\%)	    &	81 (38.76\%)	&	391 (23.78\%)\\
			Miscellaneous           & 59 (21.38\%)	&	22 (11.17\%)	&	11 (9.32\%)	        &	79 (11.29\%)	&	37 (25.69\%)	    &	25 (11.96\%)	&	233 (14.17\%)\\ \hline
			Total counter           & 276           & 197               & 118                   & \cellcolor{green}700               & 144                   & 209               &   1644 \\ \hline
			Total hate      & 120 & 110 & 43  & \cellcolor{green}143 & 91  & 99  & 606  \\

			\hline
	\end{tabular}}
	\caption{Counterspeech strategies used by various target communities. The percentage given in the bracket are normalized by the total number of counterspeech in each category. Please note that the total reported is more than the 1290 counterspeech replies because of the presence of multiple target communities in a single tweet.}
	~\label{tab:counterspeech_per_target_category}
\end{table*}

\section{Target community analysis}

Our final dataset consisted of 558 hatespeech tweets from 548 Hate Accounts (HAs)
and 1290 direct replies that were counterspeech from 1239 Counterspeech Accounts (CAs). Of the total 1711 tweets that were replies to the 558 hate tweets, 75.39\% (1290) were counterspeech tweets. This is almost double of what is reported in ~\cite{mathew2018thou}. We argue that the main reason behind this could be the public nature of Twitter as opposed to the semi-anonymous nature of Youtube. We also found that 82.09\% of the counterspeech employed only one kind of strategy. The $p$-values reported are calculated using Mann-Whitney U test.

\subsection{Strategies used by CAs}

The CAs used different strategies to tackle the hatespeech. We can observe from Table~\ref{tab:counterspeech_per_target_category} that different target communities adopt different measures to respond to the hateful tweets. The largest fraction (23.60\%) of the hateful tweets seems to be religious in nature. They also receive the highest number of counterspeech with an average of 4.9 counterspeech for every religious hateful tweet as compared to only 2.7 counterspeech per hateful tweet for the entire dataset. The religious CAs seems to be using two strategies more as compared to other target communities: presentation of facts and pointing out hypocrisy. 

In case of the nationality target communities, CAs seem to be relying more on affiliation and using images in their counter speech. The CAs for the ethnicity use warning of the consequences and denounce the hatespeech more as compared to other target communities.

The CAs for the target communities associated with sexuality rely on humor and positive tone to counter the hateful messages. In case of the hatespeech which target the physical traits, the CAs heavily use hostile language to counter them.

Irrespective of the community, the counterspeakers seem to be using hostile language a lot with the lowest being for the religion (17.71\%) and highest for the physical traits (45.14\%). We can also observe that 27.67\% of the counterspeech tweets used hostile language as a strategy. This is less than the 35\% reported in~\cite{mathew2018thou} on YouTube. One of the main reasons for this could be the more public nature of Twitter.

\subsection{Use of images/videos}
We observe that only 12 (2.15\%) of hate tweets contained images/videos whereas 145 (11.24\%) of the counter tweets had images/videos. We also look into the replies of these 12 hate tweets which use images/videos to get a better understanding. Interestingly, we found that 69 (47.59\%) out of the 145 counter tweets which use images/videos were in response to hateful tweets which themselves used images/videos. Another interesting observation was that the counterspeech involving images/videos are liked more by the Twitter community as compared to other strategies. The counter images received the highest likes among all the counterspeech strategies with a median of 2.5 likes (average = 10).












\section{User level analysis}
In this section, we characterize the HAs and CAs based on their activity and tweet history. We first collect for each user their last 3200 tweets using the Twitter API\footnote{Twitter API: \url{https://developer.twitter.com/en/docs/tweets/timelines/api-reference/get-statuses-user_timeline.html}}. This would also give us other information about the users such as the number of tweets, number of followers, friends etc.

\begin{figure*}[!ht]
	\centering
	\includegraphics[width=\textwidth]{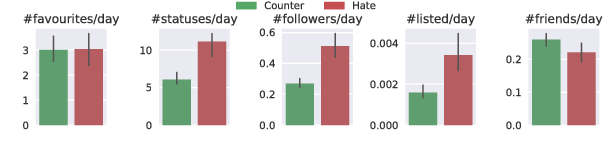} 
	\caption{ Average values for several user profile properties for Hateful and Counter user normalized by the account age. Error bars represent 95\% confidence interval.} 
	\label{fig:user_profile_property}
\end{figure*}

\subsection{User activity} We normalize the user-characteristics by dividing them by the age of the account. Thus each of the user properties are divided by the number of days since the start of the account. In Figure~\ref{fig:user_profile_property}, we can observe several striking differences in the nature of the accounts. We observe that the hate users are more ``popular'' in that they \textit{tweet more}, \textit{have more followers}, and \textit{are part of more public lists} as compared to the counter users ($p$-value\textless0.001).

The counter users have \textit{more friends} per day as compared to hate users ($p$-value\textless 0.005). 


\subsection{Creation dates} 
\begin{figure}[!ht]
	\centering
	\includegraphics[width=.4\textwidth]{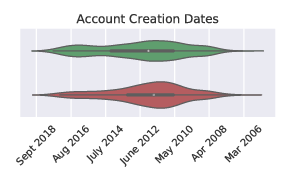} 
	\caption{KDEs of the account creation dates of Hateful and Counter users.}
	~\label{fig:account_age}    
\end{figure}

We analyze the account creation dates of hate and counter users as shown in Figure~\ref{fig:account_age}. Previous works on hate users \cite{ribeiro2017like,elsherief2018peer} have reported that the Twitter accounts of hateful users have relatively less age as compared to the normal users. Our results, in contrast, do not seem to support this observation. We find that the \textit{hate accounts are older than the counter accounts} ($p$-value\textless0.01). We argue that the main reason for this is due to the way we collected our dataset. Around 80\% of the hateful tweets are older than Dec 2017. Around this time Twitter first started enforcing stricter rules on abusive content, which means that they managed to bypass the platform's strict guidelines. Thus, many of the hate accounts are much older. 


\subsection{Suspended and deleted accounts} 
The Twitter API returns an error message when the user account is suspended or the user is not found. According to Twitter\footnote{https://help.twitter.com/en/managing-your-account/suspended-twitter-accounts}, the common reasons for account suspension include spam, risk to Twitter's security, or abusive tweet or behavior. Twitter accounts are not found when the user does not exist. This may be because the user deactivated her account or that the account was permanently deleted after thirty days of deactivation. We found that 4.56\% and 4.28\% of the hate and counter users, respectively, were deleted.

\subsection{Verified accounts}
Next, we look for the presence of verified accounts in our dataset. We found that 3.28\% of the hate users are verified; in contrast, 0.56\% of the counter users have verified accounts. We found that the hateful tweets by the HAs generated much more audience as compared to the CAs. On average (median) the verified HAs received 356.42 (57.5), 80.42 (13), 24.08 (16) likes, retweets, and replies, respectively. This is much higher than the verified CAs who received 1.8 (2.0), 0.4 (0.0), 0.4 (0.0) likes, retweets, and replies, respectively.





\begin{figure}[!t]
	\centering	
	\begin{subfigure}[t]{0.23\textwidth}
		\includegraphics[width=\textwidth]{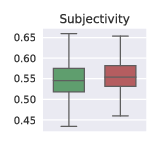}
		\label{fig:user_tweet_subjectivity}
	\end{subfigure}
	\begin{subfigure}[t]{0.23\textwidth}
		\includegraphics[width=\textwidth]{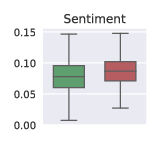}
		\label{fig:user_negative_sentiment}
	\end{subfigure}
	\begin{subfigure}[t]{0.23\textwidth}
		\includegraphics[width=\textwidth]{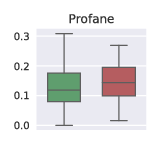}
		\label{fig:user_profanity}
	\end{subfigure}
	\begin{subfigure}[t]{0.23\textwidth}
		\includegraphics[width=\textwidth]{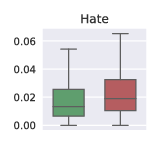}
		\label{fig:user_hate}
	\end{subfigure}
	\caption{Boxplots for the distribution of Subjectivity, Negative Sentiment, Profanity, and Hatespeech for the Hateful and Counter users} 
	\label{fig:user_text_analysis}
\end{figure}

\begin{figure*}[!ht]
	\centering	
	\begin{subfigure}[b]{0.45\textwidth}
		\includegraphics[width=\textwidth]{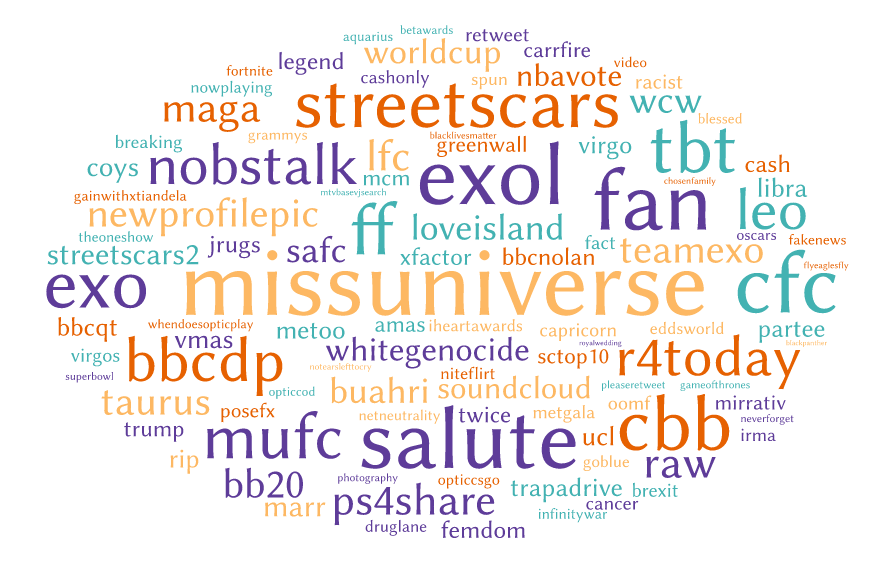}
		\caption{Hate users}
		\label{fig:hashtag_hate_wordcloud}
	\end{subfigure}
	\begin{subfigure}[b]{0.45\textwidth}
		\includegraphics[width=\textwidth]{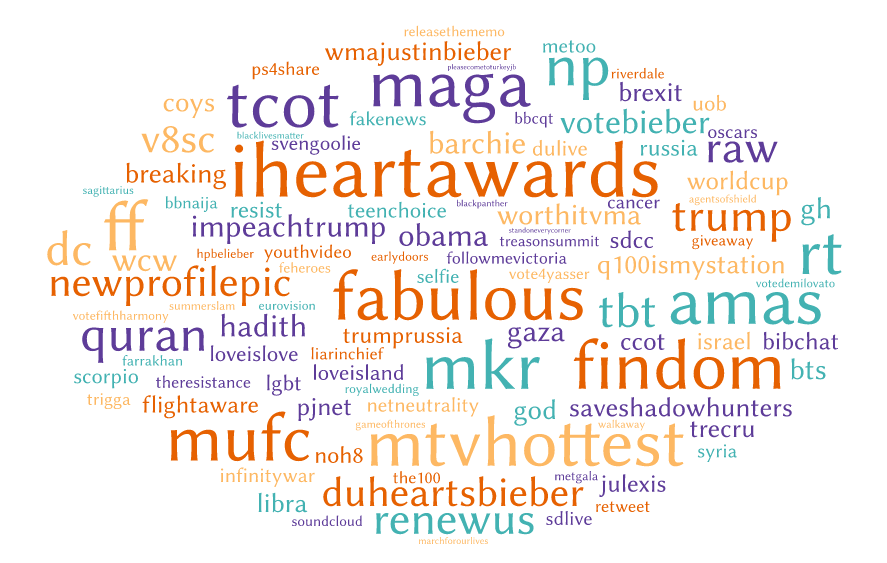}
		\caption{Counter users}
		\label{fig:hashtag_counter_wordcloud}
	\end{subfigure}
	
	\caption{Word cloud of the user hashtags. There are several hashtags such as `maga', `fakenews', `worldcup' that are common in both types of users. Notice the inclusion of hashtags such as `WhiteGenocide' in the word cloud for the hate users. The counter users seem to use the hashtags such as `loveislove' to show their support for different communities. 
	} 
	\label{fig:user_hashtag_wordcloud}
\end{figure*}

\subsection{Tweet analysis}
To understand how the different set of users express their emotions, we make use of the users' tweet history. We apply VADER~\cite{hutto2014vader} to find the average negative sentiment expressed by the users. In Figure~\ref{fig:user_text_analysis}, we observe that the tweets from hateful users  express more negative sentiments compared to counter users. We use TextBlob\footnote{https://github.com/sloria/textblob} to measure the subjectivity expressed in the tweets. As observed from Figure~\ref{fig:user_text_analysis}, hate accounts use more subjective tweets as compared to counters accounts ($p$\textless0.001).
In order to find the profanity expressed in the tweets, we use Shutterstock's ``List of Dirty, Naughty, Obscene, and Otherwise Bad Words''\footnote{https://github.com/LDNOOBW/List-of-Dirty-Naughty-Obscene-and-Otherwise-Bad-Words}. We observe that hate accounts use more profane words as compared to counter accounts ($p$-value\textless0.05).
We use the model provided by ~\cite{davidsonautomated} to check for hatespeech and abusive language in the tweet history of user. We found that the hate users seem to use more hatespeech and abusive language as compared to counter speakers.

\subsection{Hashtags}
Using the Twitter history of the users, we plot word clouds for each type of users as shown in Figure~\ref{fig:user_hashtag_wordcloud}. As we can observe, there are several hashtags such as '\#maga', '\#fakenews', '\#worldcup', '\#newprofilepic' that are common for both the types of users. The hate accounts used hashtags such as `\#whitegenocide' which is commonly used by white nationalist/supremacists. 

\begin{figure*}[!t]
	
	\includegraphics[width=\textwidth]{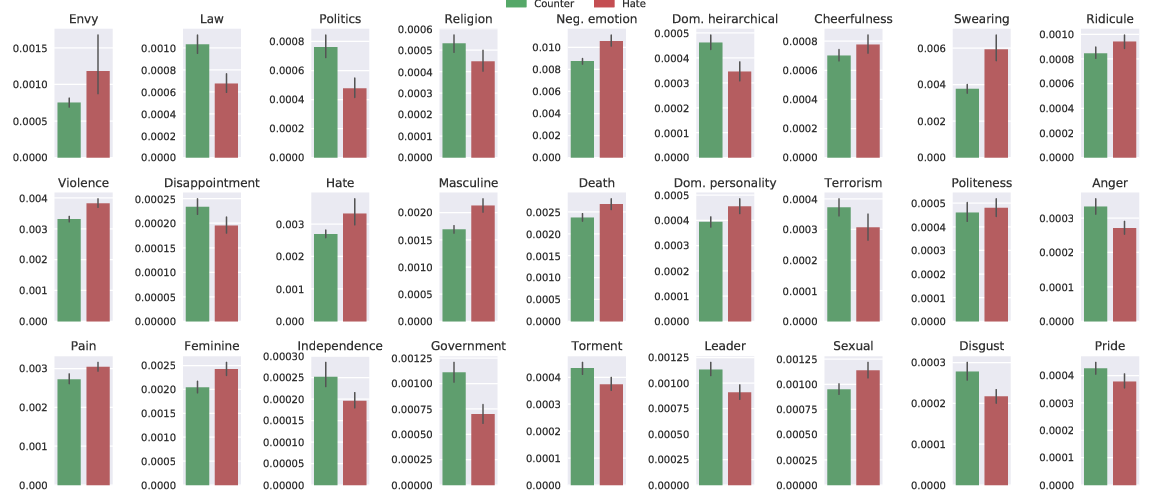} 
	\caption{Lexical analysis using Empath for the hateful and counter users. We report the median values for the relative occurrence in several categories of Empath. Error bars represent 95\% confidence intervals. That HAs score significantly higher in topics like envy, hate, negative emotion, positive emotion, ridicule, swearing terms, ugliness while CAs score significantly higher in topics like government, law, leader, pride, religion, terrorism}
	\label{fig:empath_analysis}
\end{figure*}

\subsection{Lexical analysis}
Empath~\cite{fast2016empath} is a tool that can be used to analyze text over 200 pre-built lexical categories. We characterize the users over these categories and observe several interesting results as shown in Figure~\ref{fig:empath_analysis}. The hate users seem to be using more words related to categories such as \textit{envy, hate, negative emotion, positive emotion, ridicule, swearing terms, ugliness}. These results are in alignment with the sentiment values obtained for these users. The counter users seem to be using words in the categories such as \textit{government, law, leader, pride, religion, terrorism} indicating that the counter users are more civil in nature. All the values reported have $p$-value\textless0.01.




\subsection{Personality analysis}
\begin{figure*}[!t]
	\begin{subfigure}[ht]{0.23\textwidth}
		\includegraphics[trim={3.1cm 0 2.6cm 0},clip,width=\textwidth]{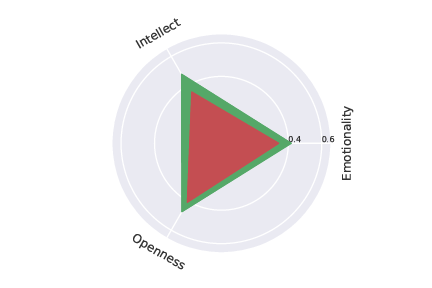}
		\caption*{Openess}
		\label{fig:IBM_openness}
	\end{subfigure}
	\begin{subfigure}[ht]{0.23\textwidth}
		\includegraphics[trim={3.0cm 0 2.6cm 0},clip,width=\textwidth]{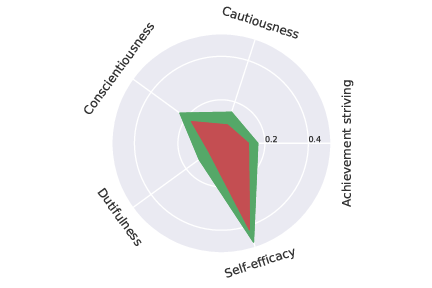}
		\caption*{Conscientiousness}
		\label{fig:IBM_Conscientiousness}
	\end{subfigure}%
	\begin{subfigure}[ht]{0.23\textwidth}
		\includegraphics[trim={2.7cm 0 2.8cm 0},clip,width=\textwidth]{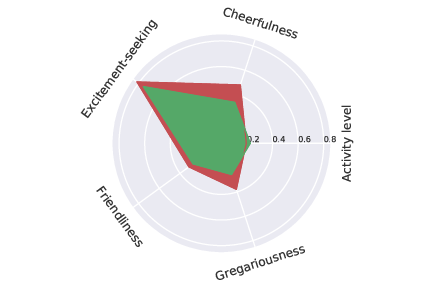}
		\caption*{Extraversion}
		\label{fig:IBM_extraversion}
	\end{subfigure}%
	\begin{subfigure}[ht]{0.23\textwidth}
		\includegraphics[trim={3.0cm 0 2.8cm 0},clip,width=\textwidth]{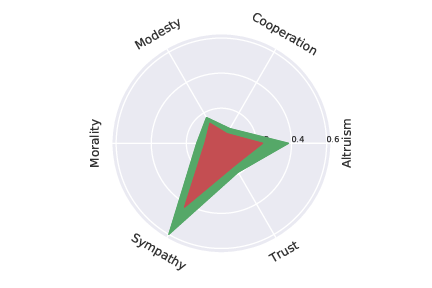}
		\caption*{Agreeableness}
		\label{fig:IBM_agreeableness}
	\end{subfigure}%

	\caption{Personality traits of the CAs and HAs. The CAs are more self-disciplined, driven, deliberate, dutiful, persistent, and self-assured while the HAs are more energetic, talkative, cheerful, excitement-seeking, outgoing, and social in nature.} 
	\label{fig:user_personality}
\end{figure*}

In order to get a better understanding of the user characteristics, we perform personality analysis on the user set. We make use of the IBM Watson Personality Insights API~\cite{Gou:2014:KSU:2556288.2557398} \footnote{IBM Personality Insights API: \url{https://www.ibm.com/watson/services/personality-insights/}} to understand the personality of the users. Previous research ~\cite{hu2016language,Liu:2017:PMS:3078714.3078733} has used this tool to understand the personality of the users. We provide all the tweets of a particular user as input to the API and the service analyzes the linguistic features to infer the Big-5 personality traits~\cite{goldberg1992development}.
We observe distinctive difference between the hate and counter users as shown in Figure~\ref{fig:user_personality}.

CAs score \textit{higher in the personality traits such as agreeableness}. They are \textit{more altruistic, modest, sympathetic}. This is intuitive since the counter speakers in many cases have a simple motive to help the hate speakers. They also seem to be \textit{more self-disciplined}. The counter speakers score higher than the hate speakers in all the sub-traits of the conscientiousness. The counter speakers are \textit{more driven, deliberate, dutiful, persistent}, and \textit{self-assured}.


The HAs seem to score \textit{higher in extraversion} indicating that they are \textit{more energetic} and \textit{talkative} in nature. They are more \textit{cheerful, excitement-seeking, outgoing}, and \textit{sociable} in nature.



\begin{figure*}[!ht]
	\centering	
	\begin{subfigure}[b]{0.45\textwidth}
		\includegraphics[width=\textwidth]{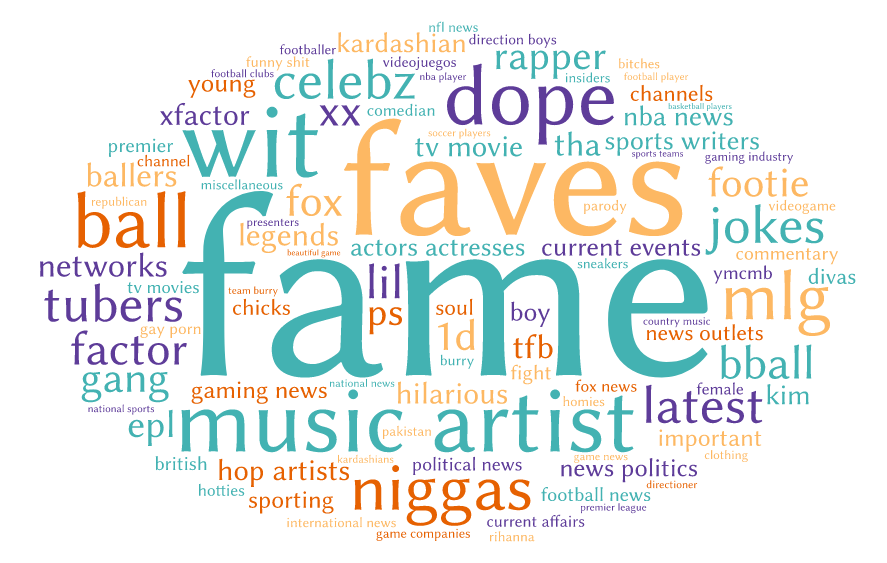}
		\caption{Hate users.}
	\end{subfigure}
	\begin{subfigure}[b]{0.45\textwidth}
		\includegraphics[width=\textwidth]{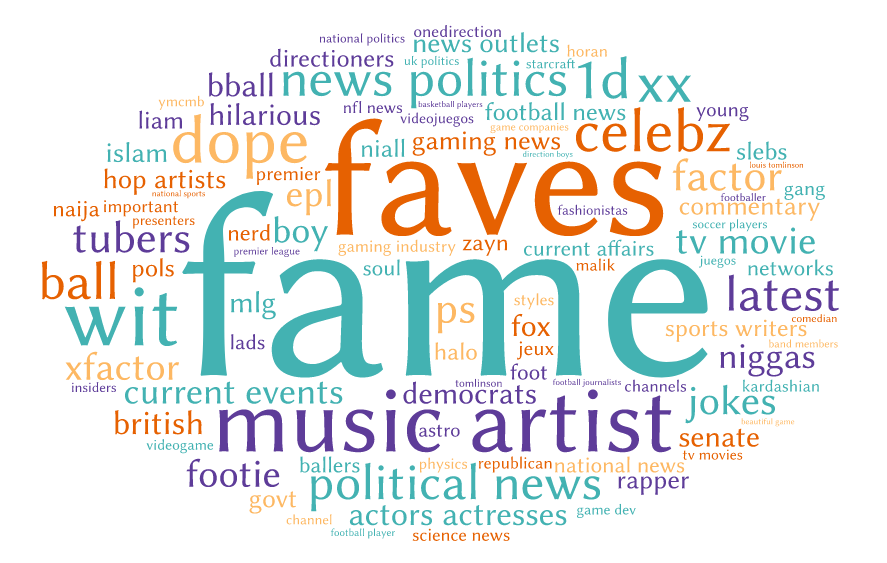}
		\caption{Counter users.}
	\end{subfigure}
	
	\caption{Word cloud of the user topical interests. Several of the topics such as `fame', `music', `artist' seems to be of interest to both hate and counter users. However while CAs subscribe to topics like `politics', `journalism' and `news' more, HAs subscribe to topics like `niggas' more.
	} 
	\label{fig:user_interest_wordcloud}
\end{figure*}
\subsection{Topical inclination of users}

We infer the topical interests of the users using the method explained in ~\cite{bhattacharya2014inferring} and utilize the `Who likes what?' web service\footnote{Topical Interest: \url{https://twitter-app.mpi-sws.org/who-likes-what}} for this task. In this method, for a given Twitter user $u$ (whose interests are to be inferred), the method first checks which other users $u$ is following, i.e., users from whom $u$ is interested in receiving information. It then identifies the topics of expertise of those users (whom $u$ is following) to infer $u$'s interests, i.e., the topics on which $u$ is interested in receiving information. Expertise is defined by the users bio or tweets via the lists feature in Twitter~\cite{ghosh2012cognos}.

In order to understand which topics are preferred by the HAs and CAs, we analyze the topical interests of the users as shown by the word cloud in Figure~\ref{fig:user_interest_wordcloud}. We can observe that hate and counter users have several topics of common interest. Topics such as `Fame', `Music', `Artist' seem to be of interest to a large proportion of hate and counter users. On a closer analysis, we found that topics such as `politics', `news', and `journalism' are preferred more by the CAs as compared to HAs ($p$-value\textless0.01). This is possibly indicative of a constructive nature of the CAs. Also the HAs tend to subscribe to topics like `niggas' significantly more.

\section{Classification}

In this section, we leverage the unique characteristics of the hate and counter user accounts to develop a predictive model. We pose this as a binary classification problem - distinguishing hate users from counter users. This early automatic distinction can help the platform to develop appropriate incentive mechanisms to demote the hate user accounts and promote the counter user accounts and also urge these users to help in overall purging of the media. ~\cite{munger2017tweetment} showed that counterspeech using automated bots can reduce instances of racist speech if the instigators are sanctioned by a high-follower white male.

\subsection{Features}
We use the following feature set for our classifier.
\begin{itemize}
	\item \textbf{TF-IDF values}: For each account, we calculate its tf-idf vector using the users tweet history.
	\item \textbf{User profile properties}: We use several user account properties such as \#favorites per day, \#tweets per day, \#followers per day, \#friends per day, \#listed per day, and whether the account is verified.
	\item \textbf{Lexical properties}: We use the vector of 200 pre-built topics as the feature set for each user.
	\item \textbf{Affect properties}: For each user account we calculate the average sentiment, profanity, and subjectivity expressed in the tweet history.
	
\end{itemize}

\subsection{Dataset} 
We first divide the dataset into training and test set in the ratio 90:10 while keeping the test set balanced. In order to find the hyperparameter of our models, we use 10\% of the training data as the validation set. We run randomized grid search to find the optimal hyperparameter values.

\subsection{Choice of classifier} 
We choose classifiers such as Support Vector Machines (SVM), Logistic Regression (LR), Random Forest (RF), Extra-Tree (ET), XGBoost (XGB), and CatBoost (CB) for this task. We use TF-IDF features along with LR as the baseline.

\subsection{Results} Table~\ref{tab:counterspeech_classification} shows the results of the classification task. We can observe that CatBoost (CB) performs the best with an accuracy of 78\% followed by XGBoost (XGB) with an accuracy of 74\%. Both the classifiers perform much better than the baseline classifier (LR + TF-IDF).

\begin{table}[!ht]
	\resizebox{\linewidth}{!}{
	\begin{tabular}{ l l l l l} 
		\hline 
		\textbf{Model} &	\textbf{Precision} &	\textbf{Recall} & \textbf{F-score} & \textbf{Accuracy}  \\ \hline \hline
		LR + TFIDF & 0.68 & 0.68 & 0.68 & 0.68 \\ \hline
		SVM &  0.64  & 0.63   & 0.62 & 0.63   \\
		LR  & 0.66   & 0.66   & 0.66  & 0.66  \\
		ET  & 0.72  & 0.70   & 0.69  & 0.70 \\
		RF &0.72 & 0.72 & 0.72 & 0.72 \\ 
		\rowcolor{green!20}XGB &  0.74   &  0.74   &  0.74 & 0.74  \\
		\rowcolor{green}CB & 0.83& 0.78& 0.77& 0.78 \\ \hline
		
	\end{tabular}}
	\caption{Evaluation results for various classifiers - Support Vector Machines (SVM), Logistic Regression (LR), Random Forest (RF), Extra-Tree (ET), XGBoost (XGB), and CatBoost (CB) on the task of classification of an account as hateful or counter. The evaluation metrics reported are calculated by taking macro average.}
	~\label{tab:counterspeech_classification}
\end{table}

\subsection{Feature ablation}
We perform feature ablation to understand the importance of the different feature types. Table~\ref{tab:feature_ablation} reports the feature ablation study carried out by CatBoost (CB) classifier. From the table, we can observe that TF-IDF feature is the most useful followed by lexical features.

\begin{table}[!ht]
	\resizebox{\linewidth}{!}{
	\begin{tabular}{ l l l l l} 
		\hline 
		\textbf{Feature excluded} &	\textbf{Precision} &	\textbf{Recall} & \textbf{F-score} & \textbf{Accuracy}  \\ \hline \hline
		\rowcolor{green}TF-IDF &  0.59   &  0.53   &  0.43 & 0.53  \\
		User profile &  0.84  & 0.79   & 0.78 & 0.79   \\
		\rowcolor{green!20}Lexical  & 0.65   & 0.56   & 0.49 & 0.56  \\
		Affect  & 0.83  & 0.77   & 0.76  & 0.77 \\ \hline
		
	\end{tabular}}
	\caption{Feature ablation study for the CatBoost (CB) classifier. The evaluation metrics reported are calculated by taking macro average.}
	~\label{tab:feature_ablation}
\end{table}

\section{Related Work}

In this section, we review some of the related literature. First, we review the works done on analysis and detection of hatespeech and harmful language. Next we summarize the literature available for countering these hateful or harmful speech.

\subsection{Hateful or harmful speech}

Hatespeech lies in a complex nexus with freedom of expression, group rights, as well as concepts of dignity, liberty, and equality~\cite{gagliardone2015countering}. Owing to this, there can be several issues in defining what constitutes as hatespeech ~\cite{benesch2014defining}. The authors usually adopt a definition that fits a general description of hatespeech. There is substantial literature on the analysis of hatespeech. 
~\cite{silva2016analyzing} studies the targets of hatespeech on Twitter and Whisper and observes that `Blacks' are among the most frequent targets of hatespeech. ~\cite{mondal2017} studies the effects of anonymity on hatespeech. There are works where the authors study the perceptions and experience of people regarding the hatespeech~\cite{leets2002experiencing,gelber2016evidencing}. In~\cite{van2015good}, the authors study the impact of hatespeech prosecution of a politician on electoral support for his party and find an immediate increase in the support for the party. 

Another line of research corresponds to detection of hatespeech in various social media platforms like Twitter~\cite{waseem2016hateful,Chatzakou:2017:MGT:3041021.3053890,davidsonautomated,Badjatiya:2017:DLH:3041021.3054223}, Facebook ~\cite{del2017hate}, Yahoo! Finance and News~\cite{Warner:2012:DHS:2390374.2390377,Djuric:2015:HSD:2740908.2742760,Nobata:2016:ALD:2872427.2883062}, and Whisper~\cite{mondal2017}. In another online effort, a Canadian NGO, the Sentinel Project\footnote{https://thesentinelproject.org/}, launched a site in 2013 called HateBase\footnote{https://www.hatebase.org/}, which invites Internet users to add to a list of slurs and insulting words in many languages.

There are some works which have tried to characterize the hateful users. In ~\cite{ICWSM1817837,ribeiro2017like}, the authors study the user characteristics of hateful accounts on twitter and found that the hateful user accounts differ significantly from normal user accounts on the basis of activity, network centrality, and the type of content they produce. In ~\cite{elsherief2018peer}, the authors perform a comparative study of the hate speech instigators and target users on twitter. They found that the hate instigators target more popular and high profile twitter users, which leads to greater online visibility. In ~\cite{hatelingo2018}, the authors focus on studying the target of the hatespeech - directed and generalized. They look into linguistic and psycholinguistic properties of these two types of hatespeech and found that while directed hate speech is more personal and directed, informal and express anger, the generalized hate is more of religious type and uses lethal words such as `murder', `exterminate', and `kill'.

\subsection{Countering hatespeech}
The current solution adopted by several organization and companies to tackle online hate speech involves blocking/suspending the account or the particular hateful post. While these options are very powerful, they tend to violate the free speech doctrine. Countespeech is considered to be a promising solution in this direction as it can help in controlling the hate speech problem and at the same time, it supports free speech. Institutions can help in educating the public about hatespeech and its implications, consequences and how to respond. Programmes such as `No Hate Speech' movement~\footref{nohatespeechmovement} and Facebook's counterspeech program~\footref{counterfb} help in raising awareness, providing support and seeking creative solutions ~\cite{citron2011intermediaries}. 

Silence in response to digital hate carries significant expressive costs as well. When powerful intermediaries rebut demeaning
stereotypes (like the Michelle Obama image) and invidious falsehoods (such as holocaust denial), they send a powerful message to readers. Because intermediaries often enjoy respect and a sense of legitimacy, using counterspeech, they can demonstrate what it means to treat others with respect and dignity~\cite{citron2011intermediaries}.

Online social media sites such as Twitter, YouTube, and Facebook have been used to study counterspeech. ~\cite{wright2017vectors} study the conversations on Twitter, and find that some arguments between strangers lead to favorable change in discourse and even in attitudes. ~\cite{ernst2017hate} study the comments in YouTube counterspeech videos related to Islam and find that they are dominated by messages that deal with devaluating prejudices and stereotypes corresponding to Muslims and/or Islam. In ~\cite{schieb2016governing}, the authors study counterspeech on Facebook and through simulation, find that the defining factors for the success of counter speech are the proportion of the hatespeech and the type of influence the counter speakers can exert on the undecided. ~\cite{Stroud2018feminist} perform case studies on feminist counterspeech. They explore the range of feminist counterspeech by engaging two representative case studies from the current online environment. ~\cite{susan2016successfullcounter} describes strategies that have favorable impact or are counterproductive on users who tweet hateful or inflammatory content. In ~\cite{mathew2018thou}, the authors curate a counterspeech dataset from YouTube and build a general classifier which achieves an F1-score of 0.73 .

This work is the first to study the interaction dynamics of the hate and counter speakers in a coupled fashion. We would also release the dataset used in our experiment as well as the model for classifying an account as hateful or counter~\footnote{Dataset and Model available here: \url{https://goo.gl/vqMPc7}}.

\section{Discussion}
Here we discuss some of the key insights that we obtain from this study. We believe that many of these could be very helpful for Twitter and, in general, other social media platforms to meaningfully combat hate content. 
\subsection{Success of counterspeech}
We found cases in which the counterspeech provided was successful in changing the mindset of the user who had posted the hateful tweet. In one case, the user who had posted the hateful message on Twitter, later apologized to everyone saying that she was really sorry for what she did. Although, we rarely receive such direct evidence of a counterspeech being successful, these cases prove that counterspeech can actually help in changing the attitude of the hate users without resorting to aggressive measures such as blocking/suspension of the account.

\subsection{New venues for hatespeech}
The harsh nature of Twitter on hatespeech has led to several user account suspension and deletion. This is true for other social media sites as well. Due to this, several new social media sites have sprung up in the recent years like Gab\footnote{https://gab.ai/home}, Wrongthink\footnote{https://wrongthink.net/} which support free speech on their site and allow users to posts contents that would not be allowed on other sites. This has resulted in gab becoming an echo chamber for right-leaning dissemination~\cite{Lima2018InsideTR}.

\subsection{Common users}
We found some users which were common in HAs and CAs. On examining these accounts we observe that these accounts take the role of hate or counter user depending on the situation. For example, we observe that a user who had performed a hate speech targeting the `LGBTQ' community also performed a counterspeech in response to a hatespeech targeting `Fat people'. We found in cases where the HAs act as counter speaker, 55\% of the times they use the hostile language strategy.




\subsection{Hate supporters}
While annotating the dataset, we also found several tweets that were in support of the hate tweet. Specifically, we annotated 223 direct replies to the hate tweets that were in support. We found several interesting properties of these hate support accounts (\textbf{HSAs}). To begin with, we found that HSAs had several similarities with the HAs. They use more profane and subjective words, their accounts are relatively newer as compared to HAs and CAs, and hashtags such as `whitegenocide' were also used by them. As per lexical analysis, the HSAs use more words in the categories such as \textit{anger, independence, suffering}. The HSAs seem to be more open in nature as compared to CAs and HAs. They are \textit{more imaginative, philosophical} and \textit{authority-challenging}.


\section{Conclusion}
In this paper, we perform the first characteristic study comparing the hateful and counterspeech accounts in Twitter. We provide a dataset of 1290 tweet-reply pairs of hatespeech and the corresponding counterspeech tweets. We observe that the counter speakers of different communities adopt different techniques to counter the hateful tweets. We perform several interesting analysis on these accounts and find that hateful accounts express more negative sentiments and are more profane. The hateful users in our dataset seem to be more popular as they tweet more and have more followers. We also find that the hate tweets by verified accounts have much more virality as compared to a tweet by a non-verified account. While the hate users seem to use words more about envy, hate, negative emotion, swearing terms, ugliness, the counter users use more words related to government, law, leader. We also build a supervised model for classifying the hateful and counterspeech accounts on Twitter and obtain an F-score of 0.77. Our work should be useful in appropriately designing incentive mechanisms to make Twitter-like platforms free of hate content.

\begin{small}
	\bibliography{Main}

\begin{thebibliography}{}

\bibitem[\protect\citeauthoryear{Badjatiya \bgroup et al\mbox.\egroup
  }{2017}]{Badjatiya:2017:DLH:3041021.3054223}
Badjatiya, P.; Gupta, S.; Gupta, M.; and Varma, V.
\newblock 2017.
\newblock Deep learning for hate speech detection in tweets.
\newblock WWW,  759--760.

\bibitem[\protect\citeauthoryear{Bartlett and
  Krasodomski-Jones}{2015}]{bartlett2015counter}
Bartlett, J., and Krasodomski-Jones, A.
\newblock 2015.
\newblock Counter-speech examining content that challenges extremism online.
\newblock {\em Demos. Available at:
  http://www.demos.co.uk/wp-content/uploads/2015/10/Counter-speech.pdf}.

\bibitem[\protect\citeauthoryear{Benesch \bgroup et al\mbox.\egroup
  }{2016a}]{susan2016successfullcounter}
Benesch, S.; Ruths, D.; Dillon, K.~P.; Saleem, H.~M.; and Wright, L.
\newblock 2016a.
\newblock Considerations for successful counterspeech.
\newblock {\em Dangerous Speech Project. Available at:
  https://dangerousspeech.org/considerations-for-successful-counterspeech/}.

\bibitem[\protect\citeauthoryear{Benesch \bgroup et al\mbox.\egroup
  }{2016b}]{susan2016counterspeech}
Benesch, S.; Ruths, D.; Dillon, K.~P.; Saleem, H.~M.; and Wright, L.
\newblock 2016b.
\newblock Counterspeech on twitter: A field study.
\newblock {\em Dangerous Speech Project. Available at:
  https://dangerousspeech.org/counterspeech-on-twitter-a-field-study/}.

\bibitem[\protect\citeauthoryear{Benesch}{2014}]{benesch2014defining}
Benesch, S.
\newblock 2014.
\newblock Defining and diminishing hate speech.
\newblock {\em Freedom from hate: State of the world’s minorities and
  indigenous peoples 2014}  18--25.

\bibitem[\protect\citeauthoryear{Bhattacharya \bgroup et al\mbox.\egroup
  }{2014}]{bhattacharya2014inferring}
Bhattacharya, P.; Zafar, M.~B.; Ganguly, N.; Ghosh, S.; and Gummadi, K.~P.
\newblock 2014.
\newblock Inferring user interests in the twitter social network.
\newblock In {\em Proceedings of the 8th ACM Conference on Recommender
  systems},  357--360.
\newblock ACM.

\bibitem[\protect\citeauthoryear{Chatzakou \bgroup et al\mbox.\egroup
  }{2017}]{Chatzakou:2017:MGT:3041021.3053890}
Chatzakou, D.; Kourtellis, N.; Blackburn, J.; De~Cristofaro, E.; Stringhini,
  G.; and Vakali, A.
\newblock 2017.
\newblock Measuring \#gamergate: A tale of hate, sexism, and bullying.
\newblock In {\em WWW '17 Companion},  1285--1290.

\bibitem[\protect\citeauthoryear{Citron and
  Norton}{2011}]{citron2011intermediaries}
Citron, D.~K., and Norton, H.
\newblock 2011.
\newblock Intermediaries and hate speech: Fostering digital citizenship for our
  information age.
\newblock {\em BUL Rev.} 91:1435.

\bibitem[\protect\citeauthoryear{Davidson \bgroup et al\mbox.\egroup
  }{2017}]{davidsonautomated}
Davidson, T.; Warmsley, D.; Macy, M.; and Weber, I.
\newblock 2017.
\newblock Automated hate speech detection and the problem of offensive
  language.

\bibitem[\protect\citeauthoryear{de Lima \bgroup et al\mbox.\egroup
  }{2018}]{Lima2018InsideTR}
de~Lima, L. R.~P.; Reis, J. C.~S.; Melo, P.~F.; Murai, F.; Silva, L.~A.;
  Vikatos, P.; and Benevenuto, F.
\newblock 2018.
\newblock Inside the right-leaning echo chambers: Characterizing gab, an
  unmoderated social system.

\bibitem[\protect\citeauthoryear{Del~Vigna \bgroup et al\mbox.\egroup
  }{2017}]{del2017hate}
Del~Vigna, F.; Cimino, A.; Dell’Orletta, F.; Petrocchi, M.; and Tesconi, M.
\newblock 2017.
\newblock Hate me, hate me not: Hate speech detection on facebook.

\bibitem[\protect\citeauthoryear{Djuric \bgroup et al\mbox.\egroup
  }{2015}]{Djuric:2015:HSD:2740908.2742760}
Djuric, N.; Zhou, J.; Morris, R.; Grbovic, M.; Radosavljevic, V.; and
  Bhamidipati, N.
\newblock 2015.
\newblock Hate speech detection with comment embeddings.
\newblock In {\em WWW '15 Companion},  29--30.

\bibitem[\protect\citeauthoryear{ElSherief \bgroup et al\mbox.\egroup
  }{2018a}]{hatelingo2018}
ElSherief, M.; Kulkarni, V.; Nguyen, D.; Wang, W.~Y.; and Belding, E.
\newblock 2018a.
\newblock Hate lingo: A target-based linguistic analysis of hate speech in
  social media.
\newblock ICWSM '18.

\bibitem[\protect\citeauthoryear{ElSherief \bgroup et al\mbox.\egroup
  }{2018b}]{elsherief2018peer}
ElSherief, M.; Nilizadeh, S.; Nguyen, D.; Vigna, G.; and Belding, E.
\newblock 2018b.
\newblock Peer to peer hate: Hate speech instigators and their targets.

\bibitem[\protect\citeauthoryear{Ernst \bgroup et al\mbox.\egroup
  }{2017}]{ernst2017hate}
Ernst, J.; Schmitt, J.~B.; Rieger, D.; Beier, A.~K.; Vorderer, P.; Bente, G.;
  and Roth, H.-J.
\newblock 2017.
\newblock Hate beneath the counter speech? a qualitative content analysis of
  user comments on youtube related to counter speech videos.
\newblock {\em Journal for Deradicalization}  1--49.

\bibitem[\protect\citeauthoryear{Fast, Chen, and
  Bernstein}{2016}]{fast2016empath}
Fast, E.; Chen, B.; and Bernstein, M.~S.
\newblock 2016.
\newblock Empath: Understanding topic signals in large-scale text.
\newblock In {\em Proceedings of the 2016 CHI Conference on Human Factors in
  Computing Systems},  4647--4657.
\newblock ACM.

\bibitem[\protect\citeauthoryear{Gagliardone \bgroup et al\mbox.\egroup
  }{2015}]{gagliardone2015countering}
Gagliardone, I.; Gal, D.; Alves, T.; and Martinez, G.
\newblock 2015.
\newblock {\em Countering online hate speech}.
\newblock UNESCO Publishing.

\bibitem[\protect\citeauthoryear{Gelber and
  McNamara}{2016}]{gelber2016evidencing}
Gelber, K., and McNamara, L.
\newblock 2016.
\newblock Evidencing the harms of hate speech.
\newblock {\em Social Identities} 22(3):324--341.

\bibitem[\protect\citeauthoryear{Ghosh \bgroup et al\mbox.\egroup
  }{2012}]{ghosh2012cognos}
Ghosh, S.; Sharma, N.; Benevenuto, F.; Ganguly, N.; and Gummadi, K.
\newblock 2012.
\newblock Cognos: crowdsourcing search for topic experts in microblogs.
\newblock In {\em Proceedings of the 35th international ACM SIGIR conference on
  Research and development in information retrieval},  575--590.
\newblock ACM.

\bibitem[\protect\citeauthoryear{Goldberg}{1992}]{goldberg1992development}
Goldberg, L.~R.
\newblock 1992.
\newblock The development of markers for the big-five factor structure.
\newblock {\em Psychological assessment} 4(1):26.

\bibitem[\protect\citeauthoryear{Gou, Zhou, and
  Yang}{2014}]{Gou:2014:KSU:2556288.2557398}
Gou, L.; Zhou, M.~X.; and Yang, H.
\newblock 2014.
\newblock Knowme and shareme: Understanding automatically discovered
  personality traits from social media and user sharing preferences.
\newblock CHI '14,  955--964.
\newblock ACM.

\bibitem[\protect\citeauthoryear{Hu \bgroup et al\mbox.\egroup
  }{2016}]{hu2016language}
Hu, T.; Xiao, H.; Luo, J.; and Nguyen, T.-v.~T.
\newblock 2016.
\newblock What the language you tweet says about your occupation.
\newblock In {\em Tenth International AAAI Conference on Web and Social Media}.

\bibitem[\protect\citeauthoryear{Hutto and Gilbert}{2014}]{hutto2014vader}
Hutto, C.~J., and Gilbert, E.
\newblock 2014.
\newblock Vader: A parsimonious rule-based model for sentiment analysis of
  social media text.
\newblock In {\em ICWSM}.

\bibitem[\protect\citeauthoryear{Leets}{2002}]{leets2002experiencing}
Leets, L.
\newblock 2002.
\newblock Experiencing hate speech: Perceptions and responses to anti-semitism
  and antigay speech.
\newblock {\em Journal of social issues} 58(2):341--361.

\bibitem[\protect\citeauthoryear{Liu \bgroup et al\mbox.\egroup
  }{2017}]{Liu:2017:PMS:3078714.3078733}
Liu, Z.; Xu, A.; Wang, Y.; Schoudt, J.; Mahmud, J.; and Akkiraju, R.
\newblock 2017.
\newblock Does personality matter?: A study of personality and situational
  effects on consumer behavior.
\newblock HT '17,  185--193.
\newblock ACM.

\bibitem[\protect\citeauthoryear{Mathew \bgroup et al\mbox.\egroup
  }{2018}]{mathew2018thou}
Mathew, B.; Tharad, H.; Rajgaria, S.; Singhania, P.; Maity, S.~K.; Goyal, P.;
  and Mukherje, A.
\newblock 2018.
\newblock Thou shalt not hate: Countering online hate speech.
\newblock {\em arXiv preprint arXiv:1808.04409}.

\bibitem[\protect\citeauthoryear{Mondal, Silva, and
  Benevenuto}{2017}]{mondal2017}
Mondal, M.; Silva, L.~A.; and Benevenuto, F.
\newblock 2017.
\newblock A measurement study of hate speech in social media.
\newblock In {\em HT}.

\bibitem[\protect\citeauthoryear{Munger}{2017}]{munger2017tweetment}
Munger, K.
\newblock 2017.
\newblock Tweetment effects on the tweeted: Experimentally reducing racist
  harassment.
\newblock {\em Political Behavior} 39(3):629--649.

\bibitem[\protect\citeauthoryear{Nobata \bgroup et al\mbox.\egroup
  }{2016}]{Nobata:2016:ALD:2872427.2883062}
Nobata, C.; Tetreault, J.; Thomas, A.; Mehdad, Y.; and Chang, Y.
\newblock 2016.
\newblock Abusive language detection in online user content.
\newblock In {\em WWW '16},  145--153.

\bibitem[\protect\citeauthoryear{Ribeiro \bgroup et al\mbox.\egroup
  }{2017}]{ribeiro2017like}
Ribeiro, M.~H.; Calais, P.~H.; Santos, Y.~A.; Almeida, V.~A.; and Meira~Jr, W.
\newblock 2017.
\newblock " like sheep among wolves": Characterizing hateful users on twitter.
\newblock In {\em WSDM workshop on Misinformation and Misbehavior Mining on the
  Web (MIS2)}.

\bibitem[\protect\citeauthoryear{Ribeiro \bgroup et al\mbox.\egroup
  }{2018}]{ICWSM1817837}
Ribeiro, M.; Calais, P.; Santos, Y.; Almeida, V.; and Jr., W.~M.
\newblock 2018.
\newblock Characterizing and detecting hateful users on twitter.

\bibitem[\protect\citeauthoryear{Schieb and Preuss}{2016}]{schieb2016governing}
Schieb, C., and Preuss, M.
\newblock 2016.
\newblock Governing hate speech by means of counterspeech on facebook.
\newblock In {\em 66th ICA Annual Conference, At Fukuoka, Japan},  1--23.

\bibitem[\protect\citeauthoryear{Silva \bgroup et al\mbox.\egroup
  }{2016}]{silva2016analyzing}
Silva, L.~A.; Mondal, M.; Correa, D.; Benevenuto, F.; and Weber, I.
\newblock 2016.
\newblock Analyzing the targets of hate in online social media.
\newblock In {\em ICWSM},  687--690.

\bibitem[\protect\citeauthoryear{Sorower}{2010}]{sorower2010literature}
Sorower, M.~S.
\newblock 2010.
\newblock A literature survey on algorithms for multi-label learning.
\newblock {\em Oregon State University, Corvallis} 18.

\bibitem[\protect\citeauthoryear{Stroud and Cox}{2018}]{Stroud2018feminist}
Stroud, S.~R., and Cox, W.
\newblock 2018.
\newblock The varieties of feminist counterspeech in the misogynistic online
  world.
\newblock In {\em Mediating Misogyny}. Springer.
\newblock  293--310.

\bibitem[\protect\citeauthoryear{Van~Spanje and De~Vreese}{2015}]{van2015good}
Van~Spanje, J., and De~Vreese, C.
\newblock 2015.
\newblock The good, the bad and the voter: The impact of hate speech
  prosecution of a politician on electoral support for his party.
\newblock {\em Party Politics} 21(1):115--130.

\bibitem[\protect\citeauthoryear{Warner and
  Hirschberg}{2012}]{Warner:2012:DHS:2390374.2390377}
Warner, W., and Hirschberg, J.
\newblock 2012.
\newblock Detecting hate speech on the world wide web.
\newblock In {\em Proceedings of the Second Workshop on Language in Social
  Media}, LSM '12,  19--26.

\bibitem[\protect\citeauthoryear{Waseem and Hovy}{2016}]{waseem2016hateful}
Waseem, Z., and Hovy, D.
\newblock 2016.
\newblock Hateful symbols or hateful people? predictive features for hate
  speech detection on twitter.
\newblock In {\em Proceedings of the NAACL student research workshop},  88--93.

\bibitem[\protect\citeauthoryear{Wright \bgroup et al\mbox.\egroup
  }{2017}]{wright2017vectors}
Wright, L.; Ruths, D.; Dillon, K.~P.; Saleem, H.~M.; and Benesch, S.
\newblock 2017.
\newblock Vectors for counterspeech on twitter.
\newblock In {\em Proceedings of the First Workshop on Abusive Language
  Online},  57--62.

\bibitem[\protect\citeauthoryear{Zubiaga \bgroup et al\mbox.\egroup
  }{2016}]{zubiaga_rumours_2016}
Zubiaga, A.; Liakata, M.; Procter, R.; Wong Sak~Hoi, G.; and Tolmie, P.
\newblock 2016.
\newblock Analysing how people orient to and spread rumours in social media by
  looking at conversational threads.
\newblock {\em PLOS ONE} 11:1--29.

\end{thebibliography}
	\bibliographystyle{aaai}
\end{small}

\end{document}